\documentclass[a4paper]{jpconf}
\usepackage{graphicx}

\def\ga{\,\,\raise0.14em\hbox{$>$}\kern-0.76em\lower0.28em\hbox
{$\sim$}\,\,}
\def\la{\,\,\raise0.14em\hbox{$<$}\kern-0.76em\lower0.28em\hbox
{$\sim$}\,\,}

\begin{document}
\title{Progress in Nuclear Astrophysics: 
a multi-disciplinary field with still many open questions 
}

\author{S Goriely, A Choplin, W Ryssens, I Kullmann}

\address{Institut d'Astronomie et d'Astrophysique, CP-226, Universit\'{e} Libre de Bruxelles, 1050 Brussels, Belgium }

\ead{Stephane.Goriely@ulb.be}

\begin{abstract}
Nuclear astrophysics is a multi-disciplinary field with a huge demand for nuclear data. Among its various fields, stellar evolution and nucleosynthesis are clearly the most closely related to nuclear physics. The need for nuclear data for astrophysics applications challenges experimental techniques as well as the robustness and predictive power of present nuclear models. Despite impressive progress for the last years, major problems and puzzles remain. In the present contribution, only a few nuclear astrophysics specific aspects are discussed. These concern some experimental progress related to the measurement of key reactions of relevance for the so-called s- and p-processes of nucleosynthesis, the theoretical effort in predicting nuclear properties of exotic neutron-rich nuclei of interest for the r-process nucleosynthesis, and the recent introduction of machine learning techniques in nuclear astrophysics applications. 
\end{abstract}

\section{Introduction}
\begin{figure}
\begin{center}
\includegraphics[scale=0.4]{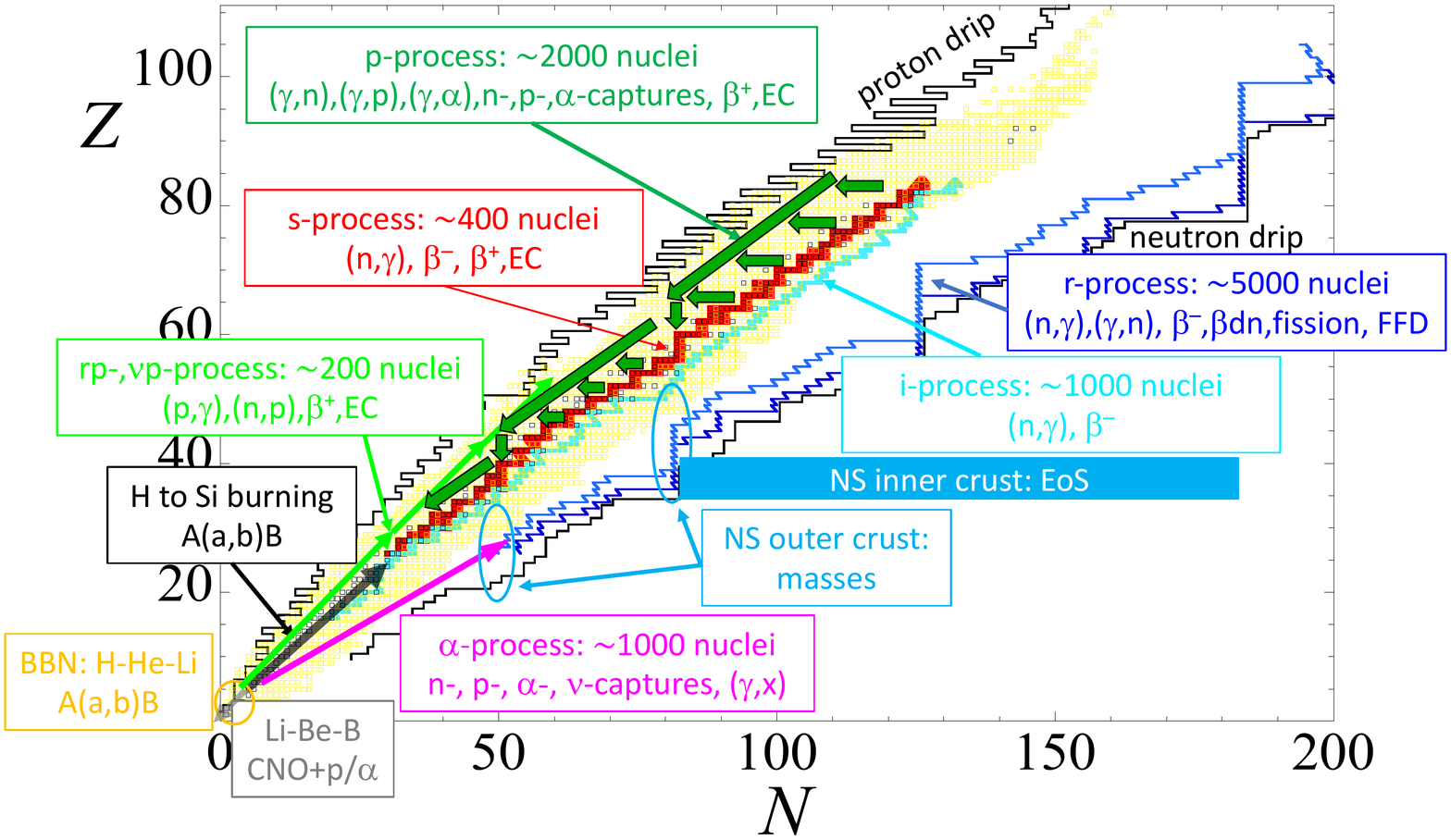}
\caption{Schematic representation in the $(N,Z)$ plane of the different astronuclear physics applications, including nucleosynthesis processes, composition and structure properties of NSs. For each process, the nuclear data needs are sketched. The open black squares correspond to stable or long-lived nuclei, the yellow squares to the nuclei for which masses have been measured and are included in the 2020 Atomic Mass Evaluation (AME) \cite{Wang21}.  Nuclei with a neutron or proton separation energies tending to zero define the neutron or proton ``drip lines'' (solid black lines), as predicted from a mass model. More details can be found in Ref.~\cite{Arnould20}.}
\label{fig_nucastro}
\end{center}
\end{figure}

The Universe is pervaded with nuclear physics imprints at all scales \cite{Arnould20,Iliadis15}. Important efforts have been devoted during the last decades to the different fields related to nucleosynthesis and stellar evolution, especially in experimental and theoretical nuclear physics, as well as in ground- or space-based astronomical observations and astrophysical modellings. In spite of many successes, major problems and puzzles remain.  In particular, experimental nuclear data only covers a minute fraction of the whole set of data required for nucleosynthesis applications. Reactions of interest often concern unstable or even exotic (neutron-rich, neutron-deficient, superheavy) species for which no experimental data exist. In addition, a large number (thousands) of unstable nuclei may be involved for which many different properties have to be determined (Fig.~\ref{fig_nucastro}). The energy range for which measurements are available is also restricted to the small range reachable by contemporary experimental setups. An additional serious difficulty comes from the fact that the nuclei are immersed in stellar environments which may have a significant impact on their static properties and the diversity and relative probabilities of their transmutation modes. The description of nuclei as individual entities has even to be replaced by the construction of an equation of state at high enough temperatures and/or densities prevailing in the cores of exploding stars and in neutron stars (NSs).  To fill the gaps, only theoretical predictions can be used. 

%****************************************************
%--------------------------------------------------

Figure~\ref{fig_nucastro} illustrates the various nuclear data needs for stellar structure, stellar evolution and nucleosynthesis applications \cite{Arnould20}. These  include the big-bang nucleosynthesis (BBN), the production of Li-Be-B by Galactic cosmic rays (in particular CNO elements interacting with protons and $\alpha$-particles), hydrostatic and explosive burning stages of stellar evolution, the rapid proton-capture process (or rp process) in X-ray bursts, the suggested $\nu$p process in exploding massive stars as well as different nucleosynthesis processes responsible for the production of elements heavier than iron, such as the slow neutron-capture process (or s process) and the intermediate neutron-capture process (or i process). The lack of nuclear data on highly neutron-deficient and neutron-rich nuclei seriously limits our ability to describe the so-called p-process in type-Ia and type-II supernovae \cite{Arnould03}, the rapid-neutron capture process (or r process) of nucleosynthesis as well as the composition of the crust of NSs \cite{Arnould07,Cowan21}. Despite the remarkable efforts of experimentalists in pushing ever closer to the neutron drip line there is unfortunately no hope of measuring the structure and interaction properties of the astrophysically relevant nuclei in the foreseeable future. For further progress one has to turn to theory. Only a few experimental and theoretical aspects are discussed in the present contribution. Readers are referred to reviews, such as Ref.~\cite{Arnould20}, for more information on the many open questions affecting nuclear astrophysics. 

\section{Recent progress on the s- and p-process nucleosynthesis}

\begin{figure}
\begin{center}
\includegraphics[scale=0.4]{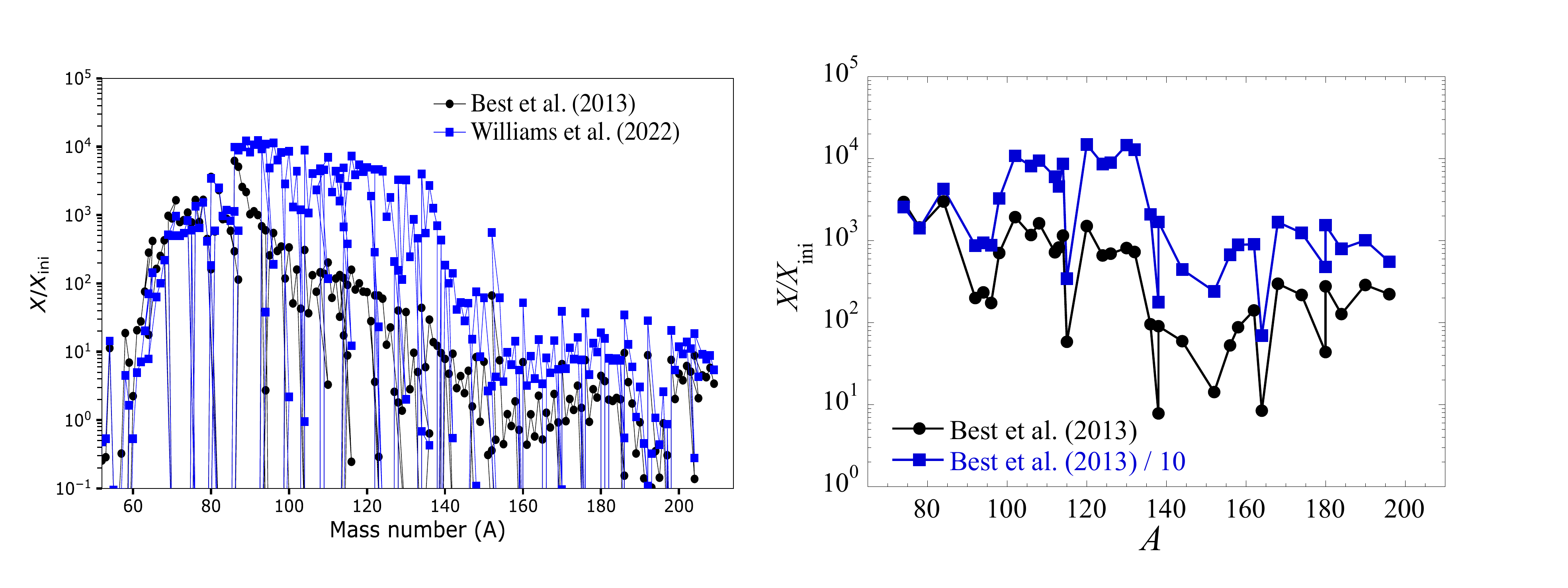}
\caption{{\it Left panel}: Overproduction factors of heavy nuclei in a rotating ($v/v_c=0.4$) low-metallicity ($Z=0.001$) massive ($M=25M_{\odot}$) star after the end of core He-burning. The s process is obtained with different predictions of the $^{17}$O+$\alpha$ rates, namely from Best et al. (2013)~\cite{Best13} or Williams et al. (2022) \cite{Williams22} and the same $^{22}$Ne+$\alpha$ rate~\cite{Longland12}. More details can be found in Ref.~\cite{Choplin18}. 
{\it Right panel}: Overproduction factors of p-nuclei produced in a rotating velocity-averaged low-metallicity ($Z=0.001$) massive ($M=25M_{\odot}$) star. Initial rotation velocities are assumed to follow the distribution observed in young B stars. More details can be found in Ref.~\cite{Choplin22}.
The p-process seeds correspond to the s process during core He-burning (right panel) with two values of the $^{17}$O($\alpha$,$\gamma$)$^{21}$Ne rates, namely the one derived in Best et al. (2013)~\cite{Best13} and the one 10 times smaller and relatively close to the rate recently derived by Williams et al.~\cite{Williams22}. }
\label{fig_sppro}
\end{center}
\end{figure}

About half of nuclides heavier than iron are produced via the  s process predominantly in the C-rich layers of asymptotic-giant-branch  (AGB) stars \cite{Karakas10,Goriely18c} as well as during core He-burning in massive stars \cite{Choplin18}. Recently, it has been shown that rotation in massive stars can significantly affect the efficiency of the s process, especially at low metallicity \cite{Choplin18,Meynet06,Frischknecht16}. 
Because of the rotational mixing operating between the H-shell and He-core during the core He-burning phase, the abundant $^{12}$C and $^{16}$O
isotopes in the convective core are mixed within the H-shell, boosting the CNO cycle and forming primary
$^{14}$N that finally leads to the synthesis of extra $^{22}$Ne, hence an increased neutron production by $^{22}$Ne($\alpha$,n)$^{25}$Mg with respect to what is found in non-rotating massive stars. The efficiency of the s process remains however sensitive to the rate of the $^{17}$O($\alpha$,$\gamma$)$^{21}$Ne reaction which, if decreased, enhances the efficiency of the neutron recycling by the competing reaction $^{17}$O($\alpha$,n)$^{20}$Ne. The updated experimental determination of the $^{17}$O($\alpha$,$\gamma$)$^{21}$Ne rate \cite{Williams22}, about 10 times smaller than the former estimate of Best et al. \cite{Best13}, is seen in Fig.~\ref{fig_sppro} to provide an efficient production of s-process nuclei beyond $A=90$ up to the second s-process peak $A\simeq 138$. Despite this important progress, the s process in massive stars remains affected by the $^{22}$Ne+$\alpha$ rates which have been recently questioned by sub-Coulomb $\alpha$-transfer measurements \cite{Jayatissa20,Adsley21}. In particular, a reduction of the $^{22}$Ne($\alpha$,n)$^{25}$Mg rate by a factor of 5 with respect to the previously determined rate of Ref.~\cite{Longland12} reduces the efficiency of the s process. Shedding light on this rate remains a priority for nucleosynthesis studies. 
 
The above-mentioned uncertainties may also impact the production of p nuclides which are known to take place during the final supernova explosion of massive stars ($M \ga 10~M_\odot$) as well as in type Ia supernovae \cite{Arnould03,Travaglio15}. In particular, the p process can develop in the O-Ne layers of the massive stars explosively heated to peak temperatures ranging between 1.7 and $ 3.3\times 10^9$~K \cite{Travaglio18}. The seeds for the p process are provided by the s process that develops before the explosion during core He-burning. In this way the O-Ne layers of solar metallicity stars that experience the p process are initially enriched in $70 \la A \la 90$ s nuclides. As discussed above, for rotating stars of sub-solar metallicity (typically around $Z=10^{-3}$), the s-process yields up to the second peak $A\simeq 138$ can be significantly increased \cite{Choplin18}, enhancing at the same time the p-process yields during the core-collapse supernova explosion \cite{Choplin22}, as illustrated in Fig.~\ref{fig_sppro}. The decrease of the $^{17}$O($\alpha$,$\gamma$)$^{21}$Ne rate \cite{Williams22} by a factor of 10 with respect to the former rate of Best et al. \cite{Best13} directly affects the synthesis of p-nuclei. A similar sensitivity is expected for the still questioned $^{22}$Ne($\alpha$,n)$^{25}$Mg rate.
%****************************************************
%--------------------------------------------------

\section{R-process nucleosynthesis}

About half of the nuclides in the valley of $\beta$ stability are known to be synthesized by the r process \cite{Arnould07,Cowan21}. This process is entirely responsible for the production of the most neutron-rich isotopes of stable elements and all long-lived elements heavier than bismuth, including Th and U in particular. Many stable nuclides are produced by both the s- and r-processes. The r process may potentially occur during core-collapse supernova explosions of massive stars %\cite{r_process_basics_4,r_process_basics_5,r_process_basics_6}  
\cite{Janka17,Wanajo18a}, in jet-like explosions of magnetorotational core-collaspe supernovae \cite{Nishimura15,Reichert21} and their collapsar remnant \cite{Siegel19b}
as well as in binary NS mergers \cite{Goriely11b,Metzger10,Just15}.
%\cite{r_process_basics_1,r_process_basics_2,r_process_basics_3} 
In such environments, large neutron densities between $10^{24}$ and $10^{34}$~cm$^{-3}$ may eventually be found leading to a series of neutron captures on timescales of the order of $\mu$s and the production of exotic neutron-rich nuclei. Temperatures above 1~GK are usually found in such astrophysical plasma, so that  photoneutron emission may slow down the nuclear flow towards the neutron drip line. On timescales of milliseconds, $\beta^-$ decay bring the flow towards heavier and heavier elements until it reaches the actinide region where fission may recycle material down to lighter mass fragments. All r-process calculations are still affected by many nuclear uncertainties in the determination of the many nuclear inputs (see Fig.~\ref{fig_nucastro}), mainly regarding exotic neutron-rich nuclei (see, e.g., \cite{Kullmann23}).

\begin{figure}
\begin{center}
\includegraphics[scale=0.33]{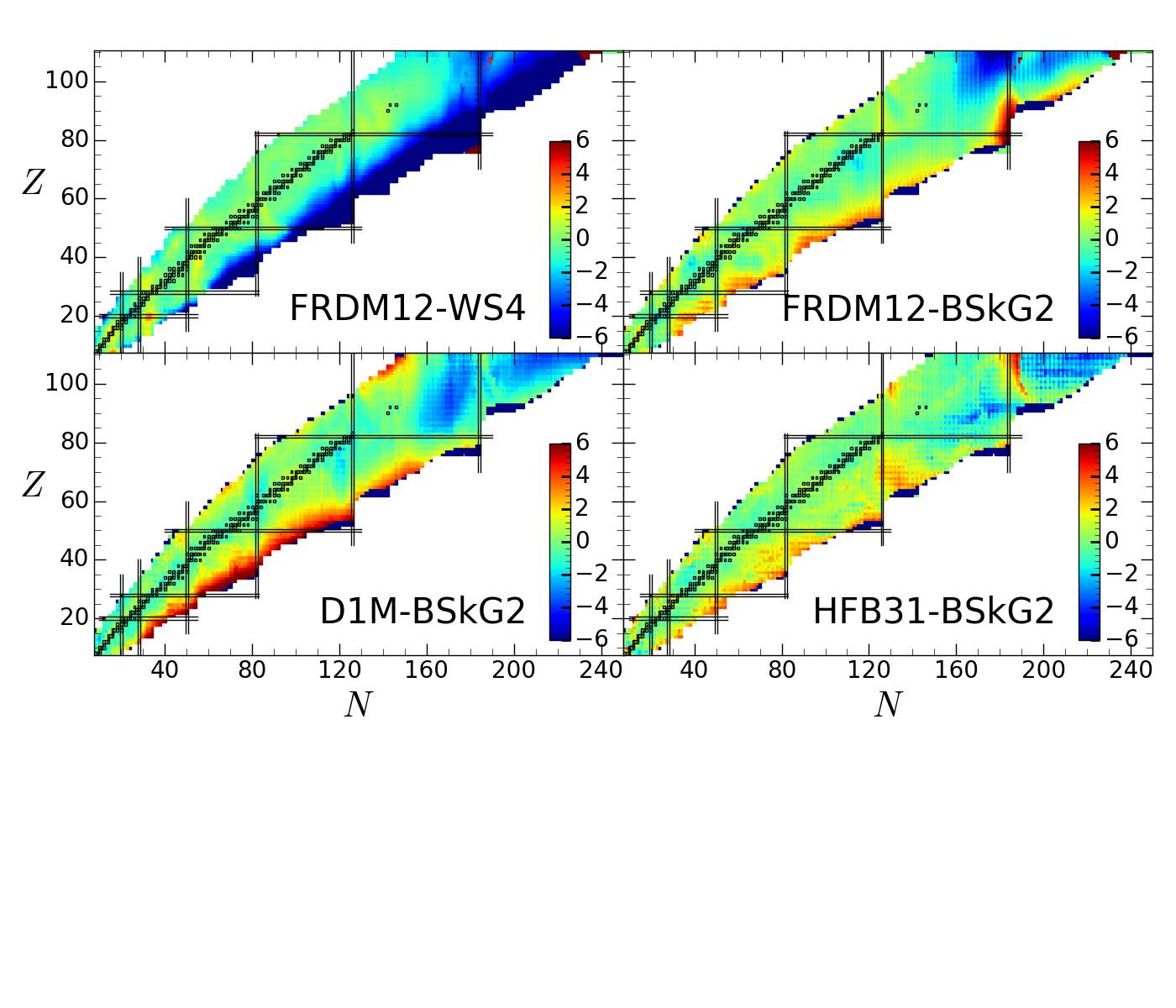}
\vskip -2.5cm
\caption{ Representation in the ($N,Z$) plane of the mass differences (in MeV) between FRDM12 \cite{Moller16}, WS4 \cite{Wang14}, HFB-31 \cite{Goriely16a},   D1M \cite{Goriely09a}, and BSkG2 \cite{Ryssens22} models for all the 8500 nuclei from $Z=8$ up to $Z=110$  between the BSkG2 proton and neutron driplines. The open squares correspond to the valley of $\beta$-stability. The double solid lines depict the neutron and proton magic numbers.}
\label{fig_dm}
\end{center}
\end{figure}

Many effective interactions within the relativistic or non-relativistic mean-field approaches have been proposed to estimate nuclear structure properties \cite{Bender03}. Only the BSk \cite{Goriely16a} and BSkG \cite{Ryssens22} effective interactions at the origin of Skyrme-HFB mass models and the D1M interaction at the origin of the Gogny-HFB mass model \cite{Goriely09a} have been fitted to the complete set of experimental masses with a root-mean-square (rms) deviations lower than 0.8~MeV and can consequently be considered for r-process applications. This contrasts with the many other Skyrme or Gogny interactions giving rise to mass predictions with an rms deviation typically larger than 2--3~MeV with respect to the bulk of known masses; {\it e.g.} masses obtained with the popular SLy4 force give an rms deviation of the order of 5~MeV \cite{Stoitsov03}. Even the UNEDF0 and UNEDF1 interactions \cite{Kortelainen10,Kortelainen12} fitted to about 72 nuclear masses end up with an rms deviation of 1.36 and 2.07~MeV on the full set of known data. With such a low accuracy, these mass models should not be used for r-process applications. Additionally, other global mass models have been developed, essentially within the macroscopic-microscopic approach, but this approach remains unstable with respect to parameter variations, as shown in the framework of the droplet model \cite{Goriely92} and as illustrated in Fig.~\ref{fig_dm} between the FRDM12 \cite{Moller16} and WS4 \cite{Wang14} mass models, especially when approaching the neutron drip line. In addition, this approach suffers from major shortcomings, such as the incoherent link between the macroscopic part and the microscopic correction or the instability of the shell correction \cite{Pearson00,Lunney03}. For this reason, more fundamental approaches, such as those based on energy density functionals are needed.

%------------------------------------------------
%--------------------------------------------------

When considering mass models obtained in relatively different frameworks, e.g the Skyrme-HFB or Gogny-HFB mass models, deviations are also found in the mass predictions away from the experimentally known region. For example, as shown in Fig.~\ref{fig_dm}, deviations up to typically $\pm 5$~MeV can be observed for exotic nuclei between HFB-31 \cite{Goriely16a}, D1M  \cite{Goriely09a} and BSkG2 \cite{Ryssens22}, especially around the $N=126$ and 184 shell closures. Neutron capture rates can consequently deviate by 3 to 5 orders of magnitude with such mass differences, essentially due to different local variations in the pairing and shell description. Such deviations by far exceeds what is acceptable for nucleosynthesis applications. For this reason, further improvements of  mass models are required. These include development of relativistic as well as non-relativistic mean field models, but also the inclusion within such approaches of the state-of-the-art beyond-mean-field corrections, like the quadrupole or octupole correlations by the Generator Coordinate Method \cite{Bender06,Robledo18} and a proper treatment of odd-$A$ and odd-odd nuclei with time-reversal symmetry breaking \cite{Ryssens22}. In addition to an accurate prediction of the known masses,  such models should also aim for an accurate description of as many other (pseudo-)observables as possible. These include charge radii and neutron skin thicknesses, fission barriers and shape isomers, spectroscopic data such as the $2^+$ energies, moments of inertia, but also infinite (neutron and symmetric) nuclear matter properties obtained from realistic calculations as well as  specific observed or empirical properties of NSs, like their maximum mass or mass-radius relations \cite{Fantina13,Pearson18}. 

\begin{figure}
\begin{center}
\includegraphics[scale=0.33]{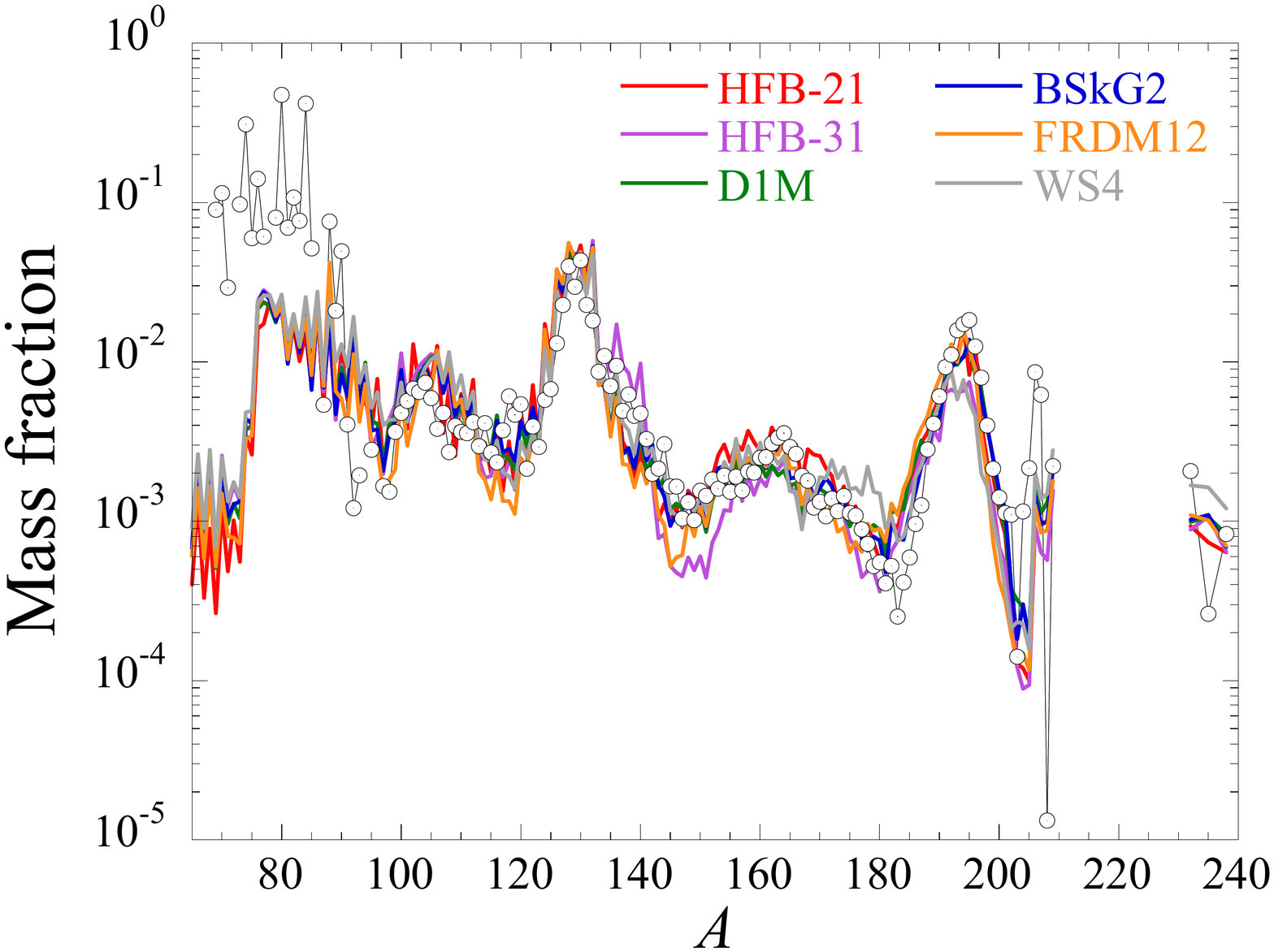}
\caption{Final mass fractions of the material ejected as a function of the atomic mass $A$ for the combined SFHo $1.35-1.35~M_{\odot}$ NS-NS merger and the M3A8m1a5 BH-torus models \cite{Just15,Kullmann23} obtained when considering, consistently, 6 different mass models in the calculation of the radiative neutron capture and photoneutron rates.
The solar system r-abundance distribution  (open circles) from \cite{Goriely99} is shown for comparison and arbitrarily normalised to the distribution obtained with HFB-21 mass model at the third r-process peak. 
}
\label{fig_nsm_mass}
\end{center}
\end{figure}

The impact of various mass models on the heavy nuclei composition of the material ejected by NS mergers is illustrated in Fig.~\ref{fig_nsm_mass} for the specific  combined SFHo $1.35-1.35~M_{\odot}$ NS-NS and the M3A8m1a5 BH-torus models \cite{Just15}. Six different mass models are considered. They all reproduce known masses with an rms deviation better than 0.8~MeV. Four are based on the HFB mean-field approach and two (FRDM12 and WS4) on the macroscopic-microscopic one. Globally, the 6 mass models are seen to give rise to abundance distributions that agree with each other and match relative well the solar system r-distribution for nuclei with $A\ga 90$. However, some local differences by a factor of 4, in either direction, can be found among them, in particular in the vicinity of $A=135-160$ and $A\simeq 200$. 
%Despite the fact that FRDM12 model does not predict any shell quenching far away from stability, no significant underabundances around $A=120$ or 180 are found, in contrast to the many claims made in the past on the basis of simple parametric r-process models \cite{Kratz14}. 
All models lead to a rather strong odd-even effects in the abundance distribution, but those are also related to the $\beta$-decay (including $\beta$-delayed) rates adopted. 
%------------------------------------------------
%--------------------------------------------------

\section{The incursion of machine learning in nuclear astrophysics}

Machine learning (ML) techniques are used more and more to support physics in estimating nuclear data that remain out of experimental reach. In particular, nuclear mass models have been complemented by ML algorithms, such as Bayesian neural networks, kernel ridge regression, Gaussian processes or radial basis functions leading to a reduction of their rms deviations to about 200~keV on all the 2500 known masses \cite{Wang14,Wu21,Shelley21}. However, improving existing mass models in this way should be done with utmost care, since ML techniques inevitably add thousands of additional parameters. These will compensate for any model defects and aid in constructing accurate interpolation, but they might also reduce the predictive power when extrapolating to unknown nuclei. 

A simple model-independent test can be performed with a neural network of about 5000 parameters, implemented via the TensorFlow algorithm ({\it https://www.tensorflow.org}) to estimate experimental masses. Considering first {\it random} sets out of the 2500 known masses \cite{Wang21}, if the algorithm is trained on 2000 masses, an rms deviation of about 0.15~MeV can be  achieved easily and validated on another 250 known masses with an rms deviation of typically 0.6~MeV. When tested on the missing 250 known masses, a typical rms deviation of 0.5~MeV can be retrieved, showing that globally our pure ML technique can reproduce the 2500 known masses with a final rms of about 0.28~MeV, even without an underlying physical model. When training now the algorithm on the same set of known masses but truncated below Pb, {\it i.e.} the 1792 known masses with $Z<82$, an rms of 0.13~MeV can easily be achieved and, as before, validated on an extra set of randomly chosen 250 masses with $Z<82$ with an rms of 0.6~MeV, both deviations being similar as in our first test. Now, if we apply our latter algorithm to the 508 known masses for $Z \ge 82$ nuclei, the rms deviation reaches a totally uncontrolled value of 37.8~MeV (with deviations up to 120~MeV for the heaviest nuclei known around $Z=110$) on those 508 masses (and about 16.9~MeV globally on all the $~2550$ masses). This simple test shows that the ML extrapolation towards the heaviest nuclei excluded from the training and validation sets is uncontrolled. Similar tests can be performed on the so-called mass residuals, {\it i.e.} the mass difference between experiment and a given physical model. In this case, the deterioration  of the mass extrapolation would obviously not be as dramatic, but would still cast doubt on the predictive power of the ML algorithm. It was indeed shown that building model biases with only $Z$ and $N$ as features leads to highly unreliable extrapolations  \cite{Perez22}. 

Despite these difficulties to control the extrapolation, ML techniques can be used successfully to enlarge the reach of existing models: either by reducing their computational requirements and/or by emulating degrees of freedom that are presently too costly. This has been illustrated in Ref.~\cite{Lasseri20} where high-quality predictions of nuclear structure properties (such as binding energies, vibrational or rotational inertias) can already be obtained with only 10\% of the full original data set. Similarly, ML algorithms can help scanning the extremely large parameter space characterising microscopic mass models and consequently improve the quality of the fit, as shown in Refs.~\cite{Ryssens22,Scamps21}.

\section{Conclusions}

Despite impressive progress for the last years, nuclear astrophysics still requires {\it (i)} dedicated experimental work on key reactions (in particular, $^{12}$C+$\alpha$, $^{12}$C+$^{12}$C, $^{22}$Ne+$\alpha$, $^{17}$O+$\alpha$, \dots), key properties (masses, charged radii, level densities, photon strength functions, optical potentials, \dots) for stable as well as unstable and exotic nuclei, and {\it (ii)} dedicated theoretical work based on models that are as ``microscopic'' as possible for experimentally inaccessible nuclei. Mean-field-based models are state-of-the-art in this respect, but they might be joined in the near future by ab-initio and shell model approaches. Support from ML will certainly be more than welcome, but mathematics will not replace physics  and special care should be given in their application to the extrapolation of data. 

\section*{Acknowledgement}
This work has been supported by the Fonds de la Recherche Scientifique (F.R.S.-FNRS; Belgium)  under Grant No IISN 4.4502.19  as well as under the EOS Project No O022818F co-funded by the Research Foundation Flanders (FWO, Belgium). The computational resources have been provided by the Consortium des Equipements de Calcul Intensif (CECI), funded by the F.R.S.-FNRS under Grant No 2.5020.11 and by the Walloon Region.

\section*{References}
\bibliographystyle{iopart-num}
\bibliography{astro}

\end{document}